\documentclass[12pt, draftcls, onecolumn]{IEEEtran}
\usepackage{amsmath}
\usepackage{amssymb}
\usepackage[dvips]{graphicx}
\usepackage{setspace}
\usepackage{epsfig}

\newcommand{\X}{\mathbf{X}}

\newcommand{\z}{\mathbf{z}}
\newcommand{\h}{\mathbf{h}}

\newcommand{\hh}{\mathbf{H}}
\newcommand{\w}{\mathbf{w}}

\newcommand{\I}{\mathbf{I}}

\newcommand{\bv}{\mathbf{v}}

\newcommand{\bu}{\mathbf{u}}

\newtheorem{lemma:bitenergylow}{Lemma}
\newtheorem{lemma:bitenergyhigh}[lemma:bitenergylow]{Lemma}

\newtheorem{prop:asympcap}{Theorem}
\newtheorem{prop:flashminbitenergy}[prop:asympcap]{Theorem}
\newtheorem{prop:flashbitenergy}[prop:asympcap]{Theorem}
\newtheorem{prop:pasympcap}[prop:asympcap]{Theorem}

\begin{document}


\title{Collaborative Relay Beamforming for Secrecy}



%
\author{\vspace{1cm}\authorblockN{Junwei Zhang and Mustafa Cenk Gursoy}
\thanks{The authors are with the Department of Electrical
Engineering, University of Nebraska-Lincoln, Lincoln, NE, 68588
(e-mails: junwei.zhang@huskers.unl.edu, gursoy@engr.unl.edu).}
\thanks{This work was supported by the National Science Foundation under Grant CCF -- 0546384 (CAREER). The material in this paper was presented in part at the 44th Annual Conference on Information Sciences and Systems (CISS), Princeton University, Princeton, NJ, in March 2010, and will be presented at the 2010 IEEE International Conference on Communications (ICC), Cape Town, South Africa, in May 2010.}}


\maketitle

\vspace{-.8cm}

\begin{spacing}{1.8}

\begin{abstract}
In this paper, collaborative use of relays to form a beamforming
system  and provide physical-layer security is investigated. In
particular,  decode-and-forward (DF)  and amplify-and-forward (AF)
relay beamforming designs under total and individual relay power
constraints are studied with the goal of maximizing the secrecy rates when perfect channel state
information (CSI) is available.
In the DF scheme, the total power constraint leads
to a closed-form solution, and in this case, the optimal beamforming structure is identified in the low and high signal-to-noise ratio (SNR) regimes.  The beamforming design under individual relay power
constraints is formulated as an optimization problem which is shown
to be easily solved using two different approaches, namely
semidefinite programming and second-order cone programming. A
simplified and suboptimal technique which reduces the computation
complexity under individual power constraints is also presented. In the
AF scheme, not having analytical solutions for the optimal beamforming design under both total and individual power constraints,
an iterative algorithm is proposed to numerically obtain the optimal beamforming structure and maximize the secrecy rates. Finally,
robust beamforming designs in the presence of imperfect CSI are investigated for DF-based relay beamforming, and optimization frameworks are provided.

\end{abstract}

\section{introduction}
The broadcast nature of wireless transmissions allows for the
signals to be received by all users within the communication range,
making wireless communications vulnerable to eavesdropping. The
problem of secure transmission in the presence of an eavesdropper
was first studied from an information-theoretic perspective in
\cite{wyner} where Wyner considered a wiretap channel model. Wyner
showed that secure communication is possible without sharing a
secret key if the eavesdropper's channel is a degraded version of
the main channel, and identified the rate-equivocation region and established the secrecy capacity of the degraded discrete memoryless wiretap channel. The
secrecy capacity is defined as the maximum achievable rate from the
transmitter to the legitimate receiver, which can be attained  while keeping the
eavesdropper completely ignorant of the transmitted messages. Later,
Wyner's result was extended to the Gaussian channel in \cite{cheong}
and recently to fading channels in \cite{Liang} and \cite{Gopala}. In addition to the
single antenna case, secrecy in multi-antenna models is addressed in
\cite{shafiee}
 and \cite{khisti}. One particular result in \cite{shafiee}
 and \cite{khisti} that is related to our study is that for the MISO secrecy channel, the optimal
 transmitting strategy is beamforming based on the generalized eigenvector of two matrices that depend on the channel coefficients.
 Regarding multiuser models, Liu  \emph{et al.} \cite{Liu} presented
inner and outer bounds on secrecy capacity regions for broadcast and
interference channels. The secrecy capacity of  the multi-antenna
broadcast channel is obtained in \cite{Liu1}.

Having multiple antennas at the transmitter and receiver has
multitude of benefits in terms of increasing the performance, and
provides the potential to improve the physical-layer security as
well. Additionally, it is well known that
even if they are equipped with single-antennas individually, users
can cooperate to form a distributed multi-antenna system by
performing relaying \cite{jnl}--\cite{kramer}. When channel side
information (CSI) is exploited, relay nodes can  collaboratively
work similarly as in a MIMO system to build a virtual beam towards
the receiver. Relay beamforming research has attracted much interest
recently (see
 e.g., \cite{Jing}--\cite{Gan1} and references therein). The optimal power
allocation at the relays has been addressed in \cite{Gan} and
\cite{Y. Jing} when instantaneous CSI is known. In \cite{luo}, the
problem of distributed beamforming in a relay network is considered
with the availability of second-order statistics  of CSI. Most recently, Zheng \emph{et
al.} \cite{Gan1} have  addressed  the robust collaborative relay
beamforming design by optimizing the weights of amplify-and-forward
(AF) relays. They maximize the worst-case signal-to-noise ratio
(SNR) assuming that CSI is imperfect but bounded. Transmit beamforming and receive beamforming strategies have been
studied extensively for over a decade. A recent tutorial
paper \cite{Gershman} provides an overview of advanced convex optimization approaches to both transmit, receive and network beamforming problems, and includes a comprehensive list of
references in this area.

Cooperative relaying under secrecy constraints was also recently studied
in \cite{dong}--\cite{aggarwal} . In \cite{dong}, a
decode-and-forward (DF) based cooperative protocol is considered,
and a beamforming system is designed for secrecy capacity
maximization or transmit power minimization. For
amplify-and-forward (AF), suboptimal closed-form solutions  that
optimize bounds on secrecy capacity are proposed in \cite{dong1}.
However, in those studies, the analysis is conducted only under
total relay power constraints and perfect CSI assumption.  In this
paper, we investigate the collaborative relay beamforming under secrecy constraints in the presence of both total and individual power
constraints with the assumptions of perfect and imperfect channel knowledge.

More specifically, our contributions in this paper are as follows:
\begin{enumerate}
\item In DF, under total power constraints, we
analytically determine the beamforming structure in the high- and low-SNR regimes.

\item In DF, under individual power constraints, not having analytical
solutions available, we provide an optimization framework to obtain the optimal beamforming that maximizes the secrecy rate. We use
the semidefinite relaxation (SDR) approach to approximate the
problem as a convex semidefinite programming (SDP) problem which can
be solved efficiently. We also provide an alternative method by
formatting the original optimization problem as a convex
second-order cone programming (SOCP) problem that can be efficiently
solved by interior point methods. Also, we describe a simplified
suboptimal beamformer design under individual power constraints.

\item In
AF, we first obtain an expression for the achievable secrecy rate, and then we show that the optimal beamforming solution that maximizes the secrecy
rate can be obtained by
semidefinite programming with a two dimensional search for both total and individual power constraints.

\item Two
robust beamforming design methods for DF relaying are described
in the case of  imperfect CSI.

\end{enumerate}

The organization of the rest of the paper is as follows. In Section
\ref{sec:DF}, we describe the channel model and study the
beamforming design for DF relaying under secrecy constraints. Beamforming for AF relaying is
investigated in Section \ref{sec:AF}. In Section \ref{sec:robust},
robust beamforming design in the case of imperfect CSI is studied.
Numerical results for the performance of different beamforming schemes are provided in
Section \ref{sec:simulation}. Finally, we conclude in
Section \ref{sec:con}.

\section{Decode-and-Forward Relaying}\label{sec:DF}

\begin{figure}
\begin{center}
\includegraphics[width = 0.6\textwidth]{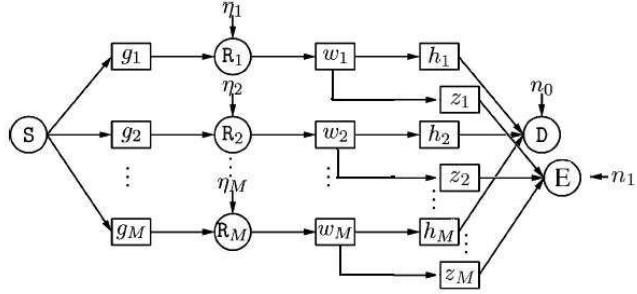}
\caption{Channel Model} \label{fig:channel}
\end{center}
\end{figure}

We consider a communication channel with a source $S$, a destination
$D$, an eavesdropper $E$, and $M$ relays $\{R_m\}_{m=1}^M$ as
depicted in Figure \ref{fig:channel}. In this model, the source $S$
tries to transmit confidential messages to destination $D$ with the help of the
relays while keeping the eavesdropper $E$ ignorant of the
information. We assume that there is no direct link between $S$ and
$D$, and $S$ and $E$.  Hence, initially messages transmitted by the
source are received only by the relays. Subsequently, relays work
synchronously and multiply the signals with
complex weights $\{w_m\}$  and produce a virtual beam point to the
destination. We denote the channel coefficient between the
source  $S$ and the $m^{th}$ relay $R_m$ as $g_m\in \mathbb{C}$, the
channel coefficient between $R_m$ and the destination $D$ as $h_m\in
\mathbb{C}$, and the channel coefficient between $R_m$ and
eavesdropper $E$ as $z_m\in \mathbb{C}$.

It is obvious that our channel is a two-hop relay network. In the
first hop, the source $S$ transmits $x_s$ to the relays with power
$E[|x_s|^2]=P_s$. The received signal at $R_m$ is given by
\begin{align}
y_{r,m}=g_m x_s+\eta_m
\end{align}
where $\eta_m$ is the background noise that has a complex, circularly symmetric Gaussian distribution with zero mean and variance of $N_m$.

In the second hop, we employ decode-and-forward transmission scheme.
In this scheme, each relay first decodes the message $x_s$ and
normalizes it as $ x_s'=x_s/\sqrt{P_s}$. Subsequently, the
normalized message is multiplied by the weight factor $w_m$ by the $m^{th}$ relay to
generate the transmitted signal $x_r=w_m x_s'$. The output power of
the $m^{th}$  relay $R_m$ is given by
\begin{align}
E[|x_r|^2]=E[|w_m  x_s'|^2]=|w_m|^2.
\end{align}
The received signals at the destination $D$ and eavesdropper $E$ are
the superpositions of the signals transmitted from the relays. These signals can be expressed, respectively, as
\begin{align}
y_d&=\sum_{m=1}^M h_m w_m  x_s' +n_0=\h^\dagger \w x_s' +n_0, \quad \text{and} \\
y_e&=\sum_{m=1}^M z_m w_m  x_s' +n_1 =\mathbf{z}^\dagger \w x_s' +n_1
\end{align}
where $n_0$ and $n_1$ are the Gaussian background noise components at $D$ and $E$, respectively, with zero mean
and variance $N_0$.  Additionally, we have defined $\mathbf{h}=[h_1^*,....h_M^*]^T,
\mathbf{z}=[z_1^*,....z_M^*]^T$, and $\w=[w_1,...w_M]^T$ where superscript $*$
denotes conjugate operation, and $(\cdot)^T$ and $(\cdot)^\dagger $ denote the
transpose and conjugate transpose, respectively, of a matrix or vector. The metrics of interest
are the received SNR levels at $D$ and $E$, which are given, respectively, by
\begin{align}
\Gamma_d=\frac{|\sum_{m=1}^M h_m w_m|^2}{N_0} \,\, \text{and} \,\,
\Gamma_e=\frac{|\sum_{m=1}^M z_m w_m|^2}{N_0}.
\end{align}
It is well-known that given the channel coefficients, the secrecy rate $R_s$ over the
channel between the relays and destination is (see e.g., \cite{cheong})
\vspace{-.5cm}
\begin{align}
R_s&=I(x_s';y_d)-I(x_s';y_e)\\
&=\log(1+\Gamma_d)-\log(1+\Gamma_e)\\
&=\log\left(\frac{N_0+|\sum_{m=1}^M h_m w_m|^2}{N_0+|\sum_{m=1}^M z_m
w_m|^2}\right) \label{eq:secrecyrate}
\end{align}
where $I(\cdot;\cdot)$ denotes the mutual information, and $x_s'$ is Gaussian distributed with zero-mean and $E[|x_s'|^2] = 1$. Coding strategies that achieve the secrecy rates involve randomization at the encoder to introduce uncertainty to the eavesdropper. Secrecy coding techniques are discussed in detail in \cite{wyner} -- \cite{Gopala}. Practical coding schemes for secure communications have been studied in \cite{Vardy} and \cite{Calderbank} for certain special cases of the wiretap channel. It is important to note that we assume in the decode-and-forward scenario that the relays use the same secrecy codebook and transmit the same signal $x_s'$ simultaneously. We further note that we throughout the text are interested in beamforming vectors that satisfy for given channel coefficients the inequality, $N_0+|\sum_{m=1}^M h_m w_m|^2 > N_0+|\sum_{m=1}^M z_m
w_m|^2$. If there are no such beamforming vectors and the ratio inside the logarithm in (\ref{eq:secrecyrate}) is less than 1, then the secrecy rate, by definition, is zero meaning that secure transmission cannot be established. The beamforming vectors which lead to zero secrecy capacity are not of interest.

In this section, we address the joint optimization of $\{w_m\}$ and hence identify the optimum
collaborative relay beamforming (CRB) direction that maximizes the secrecy rate given in (\ref{eq:secrecyrate}). Initially, we assume that the perfect knowledge of the channel coefficients is available. Later, in Section \ref{sec:robust}, we address the case in which the channel coefficients are only imperfectly known. We would like to also remark that the secrecy rate expression in (\ref{eq:secrecyrate}) in a fading environment represents the instantaneous secrecy rate for given instantaneous values of the channel fading coefficients. Hence, in such a case, our formulation considers the optimization of $\{w_m\}$ in order to maximize the instantaneous secrecy rates.

\subsection{Optimal Beamforming
under Total Power Constraints}

In this section, we consider a total relay
power constraint in the following form: $||\w||^2=\w^\dagger \w\leq P_T$. The optimization problem can now be formulated as follows:
\begin{align}\label{dftotal}
R_s(\h,\z,P_T)&=\max_{\w^\dagger \w\leq
P_T}\log\left(\frac{N_0+|\sum_{m=1}^M h_m w_m|^2}{N_0+|\sum_{m=1}^M
z_m
w_m|^2}\right) \nonumber\\
&=\log  \max_{\w^\dagger \w \leq P_T}\frac{N_0+|\sum_{m=1}^M h_m
w_m|^2}{N_0+|\sum_{m=1}^M z_m w_m|^2}\\
&=\log  \max_{\w^\dagger \w\leq P_T} \frac{\w^\dagger(\frac{N_0}{P_T}\I+\h
\h^\dagger)\w}{\w^\dagger(\frac{N_0}{P_T}\I+\z \z^\dagger)\w}\\
&=\log  \max_{\w^\dagger \w\leq P_T} \frac{\w^\dagger(N_0\I+P_T\h
\h^\dagger)\w}{\w^\dagger(N_0 \I+ P_T\z \z^\dagger)\w}\\
&=\log \lambda_{\max}(N_0\I+P_T\h \h^\dagger,N_0\I+P_T\z\z^\dagger) \label{eq:maxsecrecyrate}
\end{align}
where $\lambda_{\max}(\mathbf{A},\mathbf{B})$ is the largest
generalized eigenvalue of the matrix pair $(\mathbf{A},\mathbf{B})$
\footnote{For a Hermitian matrix $\mathbf{A} \in \mathbb{C}^{n\times
n}$ and positive definite matrix $\mathbf{B} \in \mathbb{C}^{n\times
n}$, $(\lambda,\psi)$ is referred to as a generalized eigenvalue --
eigenvector pair of $(\mathbf{A},\mathbf{B})$ if $(\lambda,\psi)$
satisfy $\mathbf{A} \psi=\lambda \mathbf{B} \psi$ \cite{matrix}.}. Hence, the maximum secrecy rate in
(\ref{eq:maxsecrecyrate}) is achieved by the optimal beamforming
vector
\begin{align}\label{woptdfto}
\w_{opt}=\varsigma \bu
\end{align}
where $\bu$ is the eigenvector that corresponds to
$\lambda_{\max}(N_0\I+P_T\h \h^\dagger,N_0\I+P_T\z\z^\dagger)$  and
$\varsigma$ is chosen to ensure $\w_{opt}^\dagger \w_{opt} =P_T$.
Note that in the first-hop of the channel model, the maximum rate we
can achieve is
\begin{align} \label{eq:firsthoprate}
R_1=\min_{m=1,\ldots,M} \log\left(1+\frac{|g_m|^2P_s}{N_m}\right).
\end{align}
Since we want all relays to successfully decode the signal transmitted from the source in the DF scenario, the rate expression in (\ref{eq:firsthoprate}) is equal to the minimum of the rates required for reliable decoding at the relays. Hence, the first-hop rate is dictated by the worst channel among the channels between the source and the relays.

The overall secrecy rate is
\begin{align}
R_{dof,s}=\min(R_1, R_s).
\end{align}
Above, we observe that having a severely weak source-relay channel can significantly degrade the performance. In these cases, other forwarding techniques (e.g., amplify-and-forward) can be preferred. Throughout the analysis of the DF scenario, we will not explicitly address these considerations and we will  concentrate on the secure communication between the relays and the destination. Hence, we will have the implicit assumption that the the source-relay links do not constitute a bottleneck for communication.

Next, we provide some remarks on the performance of collaborative
relay beamforming in the high- and low-SNR regimes. Optimal beamforming under total power constraints is studied in detail in \cite{dong} and \cite{dong1}. However, these studies have not identified the beamforming structure at low and high SNR levels. For simplicity,
we assume in the following that the noise variances at the
destination and eavesdropper are $N_0=1$.

\subsubsection{High-SNR Regime}
In the high SNR scenario, where both $P_s, P_T \to \infty  $ , we
can easily see that
\begin{align}
\lim_{P_s \to \infty} (R_1 - \log P_s) =  \min_{m=1,\ldots,M} \log
(|g_m|^2/N_m).
\end{align}
From the Corollary $4$ in Chapter $4$ of  \cite{khisti}, we can see
that
\begin{align}
\lim_{P_T \to \infty}
(R_s-\log(P_T))=\log(\max_{\tilde{\psi}}|\h^\dagger
\tilde{\psi}|^2)\label{h1}
\end{align}
where $\tilde{\psi}$ is a unit vector on the null space of
$\z^\dagger$. This result implies that choosing the
beamforming vectors to lie in the null spaces of the eavesdropper's
channel vector, i.e., having $|\sum_{m=1}^M z_m w_m|^2=\z^\dagger \w=0$, is asymptotically optimal in the high-SNR regime.
In this case, the eavesdropper cannot receive any data from the
relays, and secrecy is automatically guarantied. No secrecy coding
 is needed at the relays. This asymptotic optimality can be seen from the following discussion. Assume that we impose the constraint $\z^\dagger \w=0$. Now, the optimization problem (under the assumption $N_0 = 1$) becomes
\begin{align}
\max_{\substack{\w^\dagger \w\leq  P_T \\ \z^\dagger \w=0}}\log \left(\frac{1 + \left|\sum_{m=1}^M
h_m w_m\right|^2}{1 + \left|\sum_{m=1}^M
z_m w_m\right|^2} \right)  &= \max_{\substack{\w^\dagger \w\leq  P_T \\ \z^\dagger \w=0}}\log \left(1 + \left|\sum_{m=1}^M
h_m w_m\right|^2 \right)\label{dbrd}
\\
&= \max_{\substack{\hat{\w}^\dagger \hat{\w}\leq  1 \\ \z^\dagger \hat{\w}=0}}\log \left(1 + \left|\sum_{m=1}^M
h_m \hat{w}_m \sqrt{P_T}\right|^2 \right) \label{eq:eqvopt1}
\\
&= \log(P_T) + \max_{\substack{\hat{\w}^\dagger \hat{\w}\leq  1 \\ \z^\dagger \hat{\w}=0}}\log \left(\sqrt{\frac{1}{P_T}} + \left|\sum_{m=1}^M
h_m \hat{w}_m\right|^2 \right) \label{eq:eqvopt2}
\\
&\approx \log(P_T) + \log \left( \max_{\substack{\hat{\w}^\dagger \hat{\w}\leq  1 \\ \z^\dagger \hat{\w}=0}} \left|\sum_{m=1}^M
h_m \hat{w}_m\right|^2 \right) \label{eq:eqvopt3}
\\
&= \log( P_T)+\log(\max_{\tilde{\psi}} |\h^\dagger
\tilde{\psi}|^2)\label{h2}
\end{align}
such that $\z^\dagger \psi =0$ and $\|\psi\|^2 = 1$.
Above in (\ref{eq:eqvopt1}), we have defined $\hat{\w} = \w/\sqrt{P_T}$ for which the constraint becomes $\hat{\w}^\dagger \hat{\w}\leq  1$. The approximation in (\ref{eq:eqvopt3}) is due to the fact that $\frac{1}{\sqrt{P_T}}$ becomes negligible for large $P_T$. Hence, null space beamforming provides the same asymptotic performance as in (\ref{h1}) and is
optimal in the high-SNR regime.

Furthermore, the optimal null space beamforming
vector can be obtained explicitly.
Due to the null space constraint, we can write $\w=\hh_{z}^\bot
\bv$, where $\hh_{z}^\bot$ denotes the projection matrix onto the
null space of $\z^\dagger$. Specifically, the columns of
$\hh_{z}^\bot$ are orthonormal vectors that form the basis of the
null space of $\z^\dagger$. In our case,  $\hh_{z}^\bot$ is an
$M\times(M-1)$ matrix. The power constraint $\w^\dagger \w=
\bv^\dagger {\hh_{z}^\bot}^\dagger \hh_{z}^\bot \bv=\bv^\dagger
\bv\leq P_T$. Then, the optimization problem can be recast as
\begin{align}\label{dftotal}
\max_{\w^\dagger \w\leq  P_T}\log
\left(1+\left|\sum_{m=1}^M h_m w_m\right|^2\right)
&=\log \left(1+\max_{\w^\dagger \w\leq  P_T} (\w^\dagger \h \h^\dagger \w)\right)\\
&=\log \left(1+\max_{\bv^\dagger \bv\leq  P_T}
(\bv^\dagger{\hh_{z}^\bot}^\dagger \h \h^\dagger
{\hh_{z}^\bot}\bv)\right)\\
&=\log \left(1+P_T\lambda_{\max}
({\hh_{z}^\bot}^\dagger \h \h^\dagger {\hh_{z}^\bot})\right)\label{a}\\
&=\log \left(1+P_T \h^\dagger {\hh_{z}^\bot}{\hh_{z}^\bot}^\dagger
\h\right)\label{b}.
\end{align}
Therefore, the optimum null space beamforming vector $\w$ is
\begin{align}\label{woptdfto1}
\w_{opt,n}=\hh_{z}^\bot \bv=\varsigma_1 \hh_{z}^\bot
{\hh_{z}^\bot}^\dagger \h
\end{align}
where $\varsigma_1$ is a constant that is introduced to satisfy the power constraint.

\subsubsection{Low-SNR Regime}

In the low SNR regime, in which both $P_s, P_T \to 0$, 
we can  see that
\begin{align}
\lim_{P_s \to 0} \frac{R_1}{P_s} = \min_{m=1,\ldots,M}
\frac{|g_m|^2}{N_m}, \text{ and}
\\
\lim_{P_s \to 0} \frac{R_s}{P_T} =  \lambda_{\max}(\h \h^ \dagger-\z
\z^\dagger).
\end{align}
Thus, in the low SNR regime, the direction of the optimal beamforming vector
approaches that of the eigenvector that corresponds to the largest eigenvalue of $\h \h^ \dagger-\z \z^\dagger$. A similar result is shown in a multiple-antenna setting in \cite{gursoy}.

%

\subsection{Optimal Beamforming under Individual Power Constraints}

In a multiuser network such as the relay system we study in this
paper, it is practically more relevant to consider individual power
constraints as wireless nodes generally operate under such
limitations. Motivated by this, we now impose $|w_m|^2 \leq p_m$
 $\forall m $ or equivalently $|\w|^2 \leq \mathbf{p}$ where
$|\cdot|^2$ denotes the element-wise norm-square operation and
$\mathbf{p}$ is a column vector that contains the components
$\{p_m\}$. In what follows, the problem of interest will be again be the maximization of the secrecy rate or equivalently the
maximization of the term inside logarithm function of $R_s$
(\ref{eq:secrecyrate}) but now under individual power constraints:
\begin{align}
&\max_{|\w|^2 \leq \mathbf{p}} \frac{N_0+|\sum_{m=1}^M h_m
w_m|^2}{N_0+|\sum_{m=1}^M z_m w_m|^2} \label{optind1}\\
&=\max_{|\w|^2 \leq \mathbf{p}} \frac{N_0+\w^\dagger \h \h^\dagger
\w}{N_0+\w^\dagger \z \z^\dagger \w}. \label{optind}
\end{align}

\subsubsection{Semidefinite Relaxation (SDR) Approach} We first
consider a semidefinite programming method similar to that in
\cite{luo}. Using the definition $\X\ \triangleq \w \w ^\dagger$, we
can rewrite the optimization problem in (\ref{optind}) as
\begin{align}\label{SDR1}
\begin{split}
\max_{\X}~~ &\frac{N_0+tr(\h \h^\dagger \X)}{N_0+tr(\z \z^\dagger
\X)}
\\
s.t ~~& ~~diag(\X)\leq \mathbf{p} \\
&~~ rank ~~\X=1,~~~ and ~~~\X\succeq 0
\end{split}
\end{align}
or equivalently as
\begin{align}
\begin{split}
\max_{\X, t} &~~~t \label{SDR} \\
s.t &~~ tr(\X(\h\h^\dagger- t\z\z^\dagger))\geq N_0(t-1),\\
~~& ~~diag(\X)\leq \mathbf{p},  \\
&~~ rank ~~\X=1,~~~ and ~~~\X\succeq 0
\end{split}
\end{align}
where $tr(\cdot)$ represents the trace of a matrix, $diag(\X)$ denotes the vector whose components are the diagonal elements of $\X$,
and $\X\succeq 0$ means that $\X$  is a symmetric positive
semi-definite matrix. The optimization problem in (\ref{SDR}) is not
convex and may not be easily solved. Let us now ignore the rank
constraint in (\ref{SDR}). That is, using a semidefinite relaxation
(SDR), we aim to solve the following optimization problem:
\begin{align}
\begin{split}
&\max_{\X, t} ~~~t \label{SDR2} \\
&s.t ~~ tr(\X(\h\h^\dagger- t\z\z^\dagger))\geq N_0(t-1),\\
&and ~~diag(\X)\leq \mathbf{p}, ~~~ and ~~~\X \succeq 0.
\end{split}
\end{align}
 If the matrix $\X_{opt}$ obtained by solving the optimization problem in
(\ref{SDR2}) happens to be rank one, then its principal component
will be the optimal solution to the original problem. Note that the
optimization problem in (\ref{SDR2}) is quasiconvex. In fact, for
any value of $t$, the feasible set in (\ref{SDR2}) is convex.  Let
$t_{\max}$ be the maximum value of $t$ obtained by solving the
optimization problem  (\ref{SDR2}). If, for any given $t$, the
convex feasibility problem
\begin{align}
\begin{split}
&find~~~~\X \label{SDR3}\\
&such~~that ~ tr(\X(\h\h^\dagger- t\z\z^\dagger))\geq N_0(t-1),\\
&and ~~diag(\X)\leq \mathbf{p}, ~~~ and ~~~\X \succeq 0
\end{split}
\end{align}
is feasible, then we have $t_{\max}\geq t$. Conversely, if the
convex feasibility optimization problem (\ref{SDR3}) is not
feasible, then we conclude $t_{\max} <t$. Therefore, we can check
whether the optimal value $t_{\max}$ of the quasiconvex optimization
problem in (\ref{SDR2}) is smaller than or greater than a given
value $t$ by solving the convex feasibility problem (\ref{SDR3}). If
the convex feasibility problem (\ref{SDR3}) is feasible then we know
$t_{\max}\geq t$. If the convex feasibility problem (\ref{SDR3}) is
infeasible, then we know that  $t_{\max}< t$. Based on this
observation, we can use a simple  bisection algorithm to solve the
quasiconvex optimization problem (\ref{SDR2}) by solving a convex
feasibility problem (\ref{SDR3}) at each step. We assume that the
problem is feasible, and start with an interval $[l ,u]$ known to
contain the optimal value $t_{\max}$. We then solve the convex
feasibility problem at its midpoint $t = (l + u)/2$ to determine
whether the optimal value is larger or smaller than $t$. We update
the interval accordingly to obtain a new interval. That is, if $t$
is feasible, then we set $l = t$, otherwise, we choose $u = t$ and
solve the convex feasibility problem  again. This procedure is
repeated until the width of the interval is smaller than the given
threshold. Note that the technique of using bisection search to solve the SDP feasibility
problem is also given in \cite{lzhang}. Once the maximum feasible
value for $ t_{\max}$ is obtained, one can solve
\begin{align}
\begin{split}
&\min_{\X} ~~~tr(\X) \label{SDR4} \\
&s.t ~~ tr(\X(\h\h^\dagger- t_{\max}\z\z^\dagger))\geq N_0(t_{\max}-1),\\
&and ~~diag(\X)\leq \mathbf{p}, ~~~ and ~~~\X \succeq 0
\end{split}
\end{align}
to get the solution $\X_{opt}$. (\ref{SDR4}) is a convex problem
which can be solved efficiently using interior-point based methods.

To solve the convex feasibility problem, one can use the
well-studied interior-point based methods as well. We use the
well-developed interior point method based package SeDuMi
\cite{sedumi}, which produces a feasibility certificate if the
problem is feasible, and its popular interface  Yalmip
\cite{yalmip}. In semidefinite relaxation, the solution may not be
rank one in general. Interestingly, in our extensive simulation
results, we have never encountered a case where the solution
$\X_{opt}$ to the SDP problem has a rank higher than one. In fact,
there is  always  a rank one optimal solution for our problem as
will be explained later. Therefore, we can obtain our optimal
beamforming vector from the principal component of the optimal
solution $\X_{opt}$.
\subsubsection{Second-order Cone Program (SOCP)
Approach}

The reason that the SDR method is optimal for the above problem is
that we can reformulate it as a second order cone problem
\cite{wiesel} \cite{chen} by ignoring the phase  \textbf{}in which
we optimize $\w$ directly rather than performing the optimization
over $\X = \w \w^\dagger$. This provides us with another way of
solving the optimization. The optimization problem (\ref{optind1})
is equivalent to
\begin{align}
&\max_{\w, t} ~~~~ t \label{socp}\\
&s.t\ \frac{N_0+|\h^\dagger \w|^2}{N_0+|\z^\dagger \w|^2}\geq t \label{socpt}  \\
&and~~~~|\w|^2\leq \mathbf{p}.\nonumber
\end{align}
Note that (\ref{socpt}) can be written as
\begin{align} \label{socptequiv}
\frac{1}{t}|\h^\dagger \w|^2\geq \left\| \left(\begin{array}{ccc}
\z^\dagger \w\\
 \sqrt{\left(1-\frac{1}{t}\right) N_0} \\
\end{array}\right)\right\|^2 = |\z^\dagger \w|^2 + \left(1 - \frac{1}{t}\right)N_0.
\end{align}
where the equality on the right hand side of (\ref{socptequiv}) follows from the definition of the magnitude-square of a vector.  The equivalence of (\ref{socpt}) and (\ref{socptequiv}) can easily be seen by rearranging the terms in (\ref{socptequiv}). In the above formulation, we have implicitly assumed that $t \ge 1$. Note that this assumption does not lead to loss of generality as we are interested in cases in which $\frac{N_0+|\h^\dagger \w|^2}{N_0+|\z^\dagger \w|^2} > 1$. If this ratio is less than 1, the secrecy rate, as discussed before, is zero.

Observe that an arbitrary phase rotation can be added to the
beamforming vector without affecting the constraint in (\ref{socpt}).
Thus, $\h^\dagger \w$ can be chosen to be real without loss of
generality.  We can take the
square root of both sides of (\ref{socptequiv}). The constraint becomes a
second-order cone constraint, which is convex. The optimization
problem now becomes
\begin{align}
\begin{split}
&\max_{\w, t} ~~~~ t \label{socp1}\\
&s.t ~~\sqrt{\frac{1}{t}}\h^\dagger \w \geq \left\|
\left(\begin{array}{ccc}
\z^\dagger \w\\
 \sqrt{\left(1-\frac{1}{t} \right)N_0} \\
\end{array}\right)\right\|
~\text{and}~~|\w|^2\leq \mathbf{p}.
\end{split}
\end{align}
As described in  the SDR approach, the optimal solution of
(\ref{socp1}) can be obtained by repeatedly checking the feasibility
and using a bisection search over $t$ with the aid of interior point
methods for second order cone program. Again, we use SeduMi together
with Yalmip in our simulations. Once the maximum feasible value
$t_{\max}$ is obtained, we can then solve the following second order
cone problem (SOCP) to obtain the optimal beamforming vector:
\begin{align}
\begin{split}
&\min_{\w} ~~~~ ||\w||^2 \label{socp1_2}\\
&s.t ~~\sqrt{\frac{1}{t_{\max}}}\h^\dagger \w \geq \left\|
\left(\begin{array}{ccc}
\z^\dagger \w \\
 \sqrt{\left(1-\frac{1}{t_{\max}} \right)N_0} \\
\end{array}\right)\right\|
~\text{and}~~|\w|^2\leq \mathbf{p}.
\end{split}
\end{align}
Thus, we can get the secrecy rate $R_{s,ind}$ for the second-hop
relay beamforming system under individual power constraints
employing the above two numerical optimization methods. Then,
combined with the first-hop source relay link rate $R_1$, secrecy
rate of the decode and forward collaborative relay beamforming
system becomes $ R_{dof,ind}=\min(R_1, R_{s,ind})$.
\\
\subsubsection{Simplified Suboptimal Design}

As shown above, the design of
the beamformer under individual relay power constraints requires an
iterative procedure in which, at each step, a convex feasibility
problem is solved. We now propose a suboptimal beamforming vector that can be obtained
without significant computational complexity.
%

We choose a simplified  beamformer as $\w_{sim}=\theta \w_{opt}$
where $\w_{opt}$ is given by (\ref{woptdfto}) with
$||\w_{opt}||^2=P_T=\sum p_i$ where $p_i$ is the individual power constraint for the $i^{th}$ relay, and
 we choose
\begin{align}
\theta=\frac{1}{|w_{opt,k}|/\sqrt{p_k}}
\end{align}
where $w_{opt,k}$ and $p_k$ are the $k$th entries of $\w_{opt}$ and
$\mathbf{p}$ respectively, and we choose $k$ as
\begin{align}
k=\arg \max_{1\leq i\leq M} \frac{|w_{opt,i}|^2}{p_i}
\end{align}
Substituting this beamformer  $\mathbf{w}_{sim}$ into
(\ref{eq:secrecyrate}), we get the achievable suboptimal rate under
individual power constraints.

\section{Amplify-and-Forward Relaying}\label{sec:AF}
Another common relaying scheme in practice is amplify-and-forward
relaying. In this scenario, the received signal at the $m^{th}$ relay
$R_m$ is directly multiplied by $l_mw_m$ without decoding, and forwarded to $D$. The relay output
can be written as
\begin{align}
x_{r,m}=w_m l_m (g_m x_s+ \eta_m).
\end{align}
The scaling factor,
\begin{align}
l_m=\frac{1}{\sqrt{|g_m|^2P_s+N_m}},
\end{align}
is used to ensure $E[|x_{r,m}|^2]=|w_m|^2$. The received signals at
the destination $D$ and eavesdropper $E$ are the superposition of the messages sent by the
relays. These received signals are expressed, respectively, as
\begin{align}
y_d&=\sum_{m=1}^M h_m w_m l_m (g_m x_s +\eta_m) +n_0, ~\text{and} \\
 y_e&=\sum_{m=1}^M z_m w_m l_m (g_m x_s+\eta_m) +n_1.
\end{align}
Now, it is easy to compute  the received SNR at $D$ and $E$ as
\begin{align}
\Gamma_d&=\frac{|\sum_{m=1}^M h_m g_m l_m w_m|^2 P_s}{\sum_{m=1}^M |h_m|^2l_m^2 |w_m|^2 N_m +N_0}, ~\text{and} \\
\Gamma_e&=\frac{|\sum_{m=1}^M z_m g_m l_m w_m|^2 P_s}{\sum_{m=1}^M
|z_m|^2l_m^2 |w_m|^2 N_m +N_0}.
\end{align}
The secrecy rate is now given by
\begin{align}
R_s&=I(x_s;y_d)-I(x_s;y_e)\\
&=\log(1+\Gamma_d)-\log(1+\Gamma_e)\\
&=\log\Bigg(\frac{|\sum_{m=1}^M h_m g_m l_m w_m|^2 P_s+\sum_{m=1}^M
|h_m|^2l_m^2 |w_m|^2 N_m +N_0}{|\sum_{m=1}^M z_m g_m l_m w_m|^2
P_s+\sum_{m=1}^M |z_m|^2l_m^2 |w_m|^2 N_m +N_0}\nonumber \\
&\times \frac{\sum_{m=1}^M |z_m|^2l_m^2 |w_m|^2 N_m
+N_0}{\sum_{m=1}^M |h_m|^2l_m^2 |w_m|^2 N_m +N_0}\Bigg).
\end{align}
Again, we maximize this term by optimizing $\{w_m \}$ jointly
with the aid of perfect CSI. It is obvious that we only have to
maximize the term inside the logarithm function. Let us define
\begin{align}
\mathbf{h_g}&=[h_1^* g_1^* l_1,... ,h_M^* g_M^* l_M]^T, \\
\mathbf{h_z}&=[z_1^* g_1^* l_1,... ,z_M^* g_M^* l_M]^T, \\
\mathbf{D_h}&=\text{Diag}(|h_1|^2l_1^2N_1,...,|h_M|^2l_M^2N_M), ~\text{and}\\
\mathbf{D_z}&=\text{Diag}(|z_1|^2l_1^2N_1,...,|z_M|^2l_M^2N_M).
\end{align}
Then, the received SNR at the destination and eavesdropper can be
reformulated, respectively, as
\begin{align}
\Gamma_d&=\frac{ P_s  \w^\dagger \mathbf{h_g} \mathbf{h_g}^\dagger
\w }{\w^\dagger \mathbf{D_h} \w  +N_0}
         =\frac{ P_s  tr (\mathbf{h_g} \mathbf{h_g}^\dagger \w \w^\dagger) }{tr( \mathbf{D_h} \w\w^\dagger)
         +N_0}, ~\text{and}\\
\Gamma_e&=\frac{P_s  \w^\dagger \mathbf{h_z}\mathbf{h_z}^\dagger
\w}{\w^\dagger \mathbf{D_z} \w  +N_0} =\frac{ P_s  tr (\mathbf{h_z}
\mathbf{h_z}^\dagger \w \w^\dagger) }{tr( \mathbf{D_z} \w\w^\dagger)
+N_0}.
\end{align}
With these notations, we can write the objective function of the
optimization problem as
\begin{align}
\frac{1+\Gamma_d}{1+\Gamma_e}&=\frac{1+\frac{ P_s  \w^\dagger
\mathbf{h_g} \mathbf{h_g}^\dagger \w }{\w^\dagger \mathbf{D_h} \w
+N_0}}{1+\frac{P_s  \w^\dagger \mathbf{h_z}\mathbf{h_z}^\dagger
\w}{\w^\dagger \mathbf{D_z} \w
+N_0}}\\
&=\frac{\w^\dagger \mathbf{D_h} \w  +N_0+P_s  \w^\dagger
\mathbf{h_g} \mathbf{h_g}^\dagger \w }{\w^\dagger \mathbf{D_z} \w
+N_0+P_s  \w^\dagger \mathbf{h_z}\mathbf{h_z}^\dagger \w} \times
\frac{\w^\dagger \mathbf{D_z} \w +N_0}{\w^\dagger \mathbf{D_h} \w
+N_0}\\
&=\frac{N_0+tr((\mathbf{D_h}+P_s\mathbf{h_g} \mathbf{h_g}^\dagger)\w
\w^\dagger)}{N_0+tr((\mathbf{D_z}+P_s\mathbf{h_z}
\mathbf{h_z}^\dagger)\w \w^\dagger)} \times
\frac{N_0+tr(\mathbf{D_z} \w \w^\dagger)}{N_0+tr(\mathbf{D_h} \w
\w^\dagger)}.
\end{align}
If we denote $t_1=\frac{N_0+tr((\mathbf{D_h}+P_s\mathbf{h_g} \mathbf{h_g}^\dagger)\w
\w^\dagger)}{N_0+tr((\mathbf{D_z}+P_s\mathbf{h_z}
\mathbf{h_z}^\dagger)\w \w^\dagger)}$, $t_2=\frac{N_0+tr(\mathbf{D_z} \w \w^\dagger)}{N_0+tr(\mathbf{D_h} \w
\w^\dagger)}$,  and use
the similar SDR approach as described in the DF case, we can express
the optimization problem as
\begin{align}
\begin{split}
&\max_{\X, t_1, t_2} ~~~t_1t_2  \\
&s.t ~~ tr\left(\X\left(\mathbf{D_z}-
t_2\mathbf{D_h}\right)\right)\geq N_0(t_2-1)\\
&~~ tr\left(\X\left(\mathbf{D_h}+P_s\mathbf{h_g}
\mathbf{h_g}^\dagger- t_1\left(\mathbf{D_z}+P_s\mathbf{h_z}
\mathbf{h_z}^\dagger\right)\right)\right)\geq N_0(t_1-1)\\
&and ~~diag(\X)\leq \mathbf{p},~~(and/or~~ tr(\X) \leq P_T)~~~ and
~~~\X \succeq 0.
\end{split}
\end{align}
Notice that this formulation is applied to both total relay power
constraint and individual relay power constraint which are represented by
$tr(\X) \leq P_T$ and $diag(\X)\leq \mathbf{p}$, respectively. When
there is only total power constraint, we can easily compute the
maximum values of $t_1$ and $t_2$ separately since now we have Rayleigh
quotient problems. These maximum values are
\begin{align}
t_{1,u}&=\lambda_{max}\left(\mathbf{D_h}+\frac{N_0}{P_T}\I+P_s
\mathbf{h_g}
\mathbf{h_g}^\dagger,\mathbf{D_z}+\frac{N_0}{P_T}\I+P_s\mathbf{h_z}
\mathbf{h_z}^\dagger\right),\\
t_{2,u}&=\lambda_{max}\left(\mathbf{D_z}+\frac{N_0}{P_T}\I,\mathbf{D_h}+\frac{N_0}{P_T}\I\right).
\end{align}
When there are individual power constraints imposed on the relays, we can use the bisection
algorithm similarly as in the DF case to get the maximum values
$t_{1,i,u}$ and $t_{2,i,u}$ \footnote{Subscripts $i$ in $t_{1,i,u}$ and $t_{2,i,u}$ are used to denote that these are the maximum values in the presence of individual power constraints.} for $t_1$ and $ t_2$ by repeatedly
solving the following two feasibility problems:

\begin{align}
\begin{split}
&find ~~\X  \\
&s.t~~ tr\left(\X\left(\mathbf{D_h}+P_s\mathbf{h_g}
\mathbf{h_g}^\dagger- t_1\left(\mathbf{D_z}+P_s\mathbf{h_z}
\mathbf{h_z}^\dagger\right)\right)\right)\geq N_0(t_1-1)\\
&and ~~diag(\X)\leq \mathbf{p},~~~ and ~~~\X \succeq 0,
\end{split}
\intertext{and}
\begin{split}
&find ~~\X  \\
&s.t ~~ tr\left(\X\left(\mathbf{D_z}-
t_2\mathbf{D_h}\right)\right)\geq N_0(t_2-1)\\
&and ~~diag(\X)\leq \mathbf{p},~~~ and ~~~\X \succeq 0.
\end{split}
\end{align}

Note that for both total and individual power constraints,  the maximum values of $t_1$ and $t_2$ are obtained separately above, and these values are in general attained by different $\X = \w \w^\dagger$. Now, the following strategy can be used to obtain achievable secrecy rates. For those $\X$
values that correspond to $t_{1,i,u} $ and $t_{1,u}$ (i.e., the maximum $t_1$ values under individual and total power constraints, respectively), we can compute the corresponding $t_2=\frac{N_0+tr(\mathbf{D_z} \w \w^\dagger)}{N_0+tr(\mathbf{D_h} \w
\w^\dagger)}$  and denote
them as $t_{2,i,l}$ and $t_{2,l}$ for individual
and total power constraints, respectively. Then, $\log
(t_{1,i,u}t_{2,i,l})$ and $\log (t_{1,u}$$t_{2,l})$ will serve as
our amplify-and-forward achievable rates for individual and total power constraints, respectively. With the
achievable rates, we propose the following algorithm to iteratively
search over $t_1$ and $ t_2$ to get the optimal $t_{1,o}$ and
$t_{2,o}$ that maximize the product $t_1 t_2$ by checking following
feasibility problem.
\begin {align}\label{feasible1}
\begin{split}
& find ~~~~\X\succeq 0  \\
&s.t ~  tr\left(\X\left(\mathbf{D_z}-
t_2\mathbf{D_h}\right)\right)\geq N_0(t_2-1)\\
&~~ tr\left(\X\left(\mathbf{D_h}+P_s\mathbf{h_g}
\mathbf{h_g}^\dagger- t_1\left(\mathbf{D_z}+P_s\mathbf{h_z}
\mathbf{h_z}^\dagger\right)\right)\right)\geq N_0(t_1-1)\\
&and   ~~tr(\X) \leq P_T~~~\text{if there is  total power constraint}, \\
 &or~~diag(\X)\leq \mathbf{p} ~~\text{if there is  individual power constraint}.
\end{split}
\end{align}
\subsection{Proposed Algorithm}
Define the resolution $\Delta t=\frac{t_{1,u}}{N}$ or $\Delta t=\frac{t_{1,i,u}}{N}$ for some  large $N$ for total and individual power constraints, respectively. \\
\begin{enumerate}
\item  Initialize $t_{1,o}=t_{1,u}$ , $t_{2,o}=t_{2,l}$
when total power constraint is imposed,  and $t_{1,o}=t_{1,i,u}$,
$t_{2,o}=t_{2,i,l}$ when individual power constraint is imposed.
Initialize the iteration index $i=N$.
\item Set $t_1=i\Delta t$.   If $t_1t_{2,u}<t_{1,o}t_{2,o}$
(total power constraint) or $t_1t_{2,i,u}<t_{1,o}t_{2,o}$
(individual power constraint), then go to Step (3). Otherwise,
\begin{enumerate}
\item  Let  $ t_2=\frac{t_{1,o}t_{2,o}}{t_1}$. Check the
feasibility problem (\ref{feasible1}). If it is infeasible, $i=i-1$ go to step
(2). If it is feasible, use the bisection algorithm in (\ref{feasible1}) with $t_1$ to get the maximum possible values of $t_2$
and denote this maximum as $t_{2,m}$. The initial interval in the above bisection algorithm can be chosen as
$[\frac{t_{1,o}t_{2,o}}{t_1}, t_{2,u}]$ or
$[\frac{t_{1,o}t_{2,o}}{t_1}, t_{2,i,u}]$ depending on the  power
constraints.
\item  Update $t_{1,o}=t_1$, $t_{2,o}=t_{2,m}$ ,
$i=i-1$. Go back to step (2).
\end{enumerate}
\item  Solve the following problem to get the optimal $\X$
\begin{align}
\begin{split}
& \min_{\X} ~~~~tr(\X)  \\
&s.t ~  tr\left(\X\left(\mathbf{D_z}-
t_{2,o}\mathbf{D_h}\right)\right)\geq N_0(t_{2,o}-1)\\
&~~ tr\left(\X\left(\mathbf{D_h}+P_s\mathbf{h_g}
\mathbf{h_g}^\dagger- t_{1,o}\left(\mathbf{D_z}+P_s\mathbf{h_z}
\mathbf{h_z}^\dagger\right)\right)\right)\geq N_0(t_{1,o}-1)\\
&~~\X\succeq 0 \,\,\,and
\\
&~~tr(\X) \leq P_T~~~\text{if there is  total power constraint}, \\
 &~~diag(\X)\leq \mathbf{p} ~~\text{if there is  individual power constraint}.
\end{split}
\end{align}
\end{enumerate}

\subsection{Discussion of the Algorithm}
Our algorithm is a two-dimensional search over all possible  pairs
$(t_1, t_2)$, which can produce the greatest product $t_1 t_2$,
whose logarithm will be the global maximum value of the secrecy rate. In the
following, we will illustrate how our algorithm works for
individual power constraints. Similar discussion applies to the total power
constraint case as well. The algorithm initiates with the achievable pair
$(t_{1,i,u}, t_{2,i,l})$, in which $t_{1,i,u}$ is the maximum
feasible value for $t_1$. Thus, all $t_1$ values in our search lie
in $[0, t_{1,i,u}]$. We chose the resolution parameter $N$ to
equally pick $N$ points in this interval. We then use a brute force strategy to
check each point iteratively starting from $t_{1,i,u}$ down to $0$.
In each iteration, the feasibility problem (\ref{feasible1}) is
quasi-convex. Thus, we can use the bisection search over $t_2$ to
get the greatest value of $t_2$. Note that our initial bisection interval
for $t_2$ is $[\frac{t_{1,o}t_{2,o}}{t_1}, t_{2,i,u}]$ where $t_{2,i,u}$
is the maximum feasible value for $t_2$, and
$\frac{t_{1,o}t_{2,o}}{t_1}$ is chosen so that the optimal $t_2$ we
find at the end of the bisection search will produce a product $t_1t_2$ that is greater than our
currently saved optimal $t_{1,o}t_{2,o}$. With this approach, after each
iteration, if a $t_2$ value is found, the new optimal $t_{1,o}t_{2,o}$
will be greater than the previous one.  Note that our iteration's stop
criterion is $t_1t_{2,i,u}<t_{1,o}t_{2,o}$. This means that further
decrease in the value of $t_1$ will not produce a product $t_1t_2$ that is greater
than our current $t_{1,o}t_{2,o}$. Thus, the value $t_{1,o}t_{2,o}$
at the end of this algorithm will be the global maximum since we have
already checked all possible pairs $t_1, t_2$ that are
candidates for the optimal value.

\section{Robust Beamforming Design}\label{sec:robust}

All of the beamforming methods discussed heretofore rely on the
assumption that the exact knowledge of the channel state information
is available for design. However, when the exact CSI is unavailable, the
performance of these beamforming techniques may degrade severely. Motivated by this, the problem of robust beamforming design is addressed in \cite{Bengtsson} and \cite{Chalise}.
The robust beamforming for MISO secrecy communications was  studied
in \cite{lzhang1} where the duality between
the cognitive radio MISO channel and secrecy MISO channel is exploited to transform
the robust design of the transmission strategy over the secrecy channel into a robust cognitive radio
beamforming design problem.

We additionally remark that, beside the assumption of perfect channel state
information, our previous analysis is applicable only when the
relays are fully synchronized at the symbol level. When the time
synchronization between the relays is poor, the signal replicas
passed through different relays will arrive to the destination node
with different delays. This will result in inter-symbol-interference
(ISI). To combat such ISI, the authors of \cite{Abdel} view an
asynchronous flat-fading relay network as an artificial multipath
channel (where each channel path corresponds to one particular
relay), and use the orthogonal frequency division multiplexing
(OFDM) scheme at the source and destination nodes to deal with this
artificial multipath channel. In \cite{chen1}, a filter-and-forward
protocol has been introduced for frequency selective relay networks,
and several related network beamforming techniques have been
developed. In these techniques, the relays deploy finite impulse
response (FIR) filters to compensate for the effect of
source-to-relay and relay-to-destination channels. Since the
relay synchronization problem is out of the scope of this paper, we will mainly focus on combatting the effect of
imperfect channel state information in the following discussion.

Systems robust against channel mismatches can be obtained by two
approaches. In most of robust beamforming methods, the perturbation
is modeled as a deterministic one with bounded norm which lead to a
worst cast optimization.  The other approach applied to the case in
which the CSI error is unbounded is the statistical approach which
provides the robustness in the form of confidence level measured by
probability.

Let us consider the DF case. We define
$\hat{\hh}=\hat{\h}\hat{\h}^\dagger$ and
$\hat{\mathbf{Z}}=\hat{\z}\hat{\z}^\dagger$ as the channel
estimators, and $\tilde{\hh}=\hh-\hat{\hh}$ and
$\tilde{\mathbf{Z}}=\mathbf{Z}-\hat{\mathbf{Z}}$ as the estimation
errors. First, consider the worst case optimization. In the worst case
assumption, $\tilde{\hh}$ and $\tilde{\mathbf{Z}}$ are bounded in
their Frobenius norm as $||\tilde{\hh}|| \leq \epsilon_H$,
$||\tilde{\mathbf{Z}}|| \leq \epsilon_Z$, where $\epsilon_H,
\epsilon_Z $ are assumed to be upper bounds of the channel
uncertainty. Based on the result of \cite{Bengtsson}, the robust
counterpart of previously discussed SDR-based optimization problem
can be written as
\begin{align}
\begin{split}
&\max_{\X, t} ~~~t \label{DFworst} \\
&s.t ~~ tr(\X((\hat{\hh}-\epsilon_H\I)- t(\hat{\mathbf{Z}}+\epsilon_Z \I))\geq N_0(t-1),\\
&and ~~diag(\X)\leq \mathbf{p}, ~~~ and ~~~\X \succeq 0.
\end{split}
\end{align}
Note that the total power constraint $tr(\X)\leq P_T$ can be added
into the formulation or substituted for the individual power constraint in (\ref{DFworst}).
This problem can be solved the same way as discussed before.

However, the worst-case approach requires the norms to be bounded,
which is usually not satisfied in practice. Also, this approach is
too pessimistic since the probability of the worst-case may be
extremely low. Hence,  statistical approach is a good alternative in
certain scenarios. In our case, we require the probability of the
non-outage for secrecy transmission is greater than the predefined
threshold $\varepsilon$ by imposing
\begin{gather}
Pr\left( \frac{N_0 + tr((\hat{\hh}+\tilde{\hh})\X)}{N_0 + tr((\hat{\mathbf{Z}}+\tilde{\mathbf{Z}})\X)} \geq t\right) = Pr\left(tr\left(\X(\hat{\hh}+\tilde{\hh}- t(\hat{\mathbf{Z}}+\tilde{\mathbf{Z}}))\geq N_0(t-1)\right)\right)\geq \varepsilon.
\end{gather}
Now, the optimization problem under imperfect
CSI can be expressed as
\begin{align}
\begin{split}
&\max_{\X, t} ~~~t \label{robust df} \\
&s.t ~~ Pr\left(tr\left(\X(\hat{\hh}+\tilde{\hh}- t(\hat{\mathbf{Z}}+\tilde{\mathbf{Z}}))\geq N_0(t-1)\right)\right)\geq \varepsilon,\\
&and ~~diag(\X)\leq \mathbf{p} \,\,( or ~~tr(\X)\leq P_T), ~~~ and ~~~\X
\succeq 0.
\end{split}
\end{align}
If relays are under individual power constraints, we use
$diag(\X)\leq \mathbf{p}$. Otherwise, for the case of total power constraint, we
use $tr(\X)\leq P_T$. We can also impose both constraints in the optimization.

Note that the distribution of the components of the error matrices $\tilde{\hh}$ and $\tilde{\mathbf{Z}}$ depend on the channel estimation technique and distribution of the channel coefficients. In order to simplify the analysis and provide an analytically and numerically tractable approach, we assume that the components of the
Hermitian channel estimation error matrices $\tilde{\hh}$ and
$\tilde{\mathbf{Z}}$ are independent, zero-mean, circularly
symmetric, complex Gaussian random variables with variances
$\sigma_{\tilde{H}}^2$ and $\sigma_{\tilde{Z}}^2$. Such an assumption is also used in \cite{Chalise}. Now, we can rearrange the probability
in the constraint as
\begin{align}
Pr\left(tr\left((\hat{\hh}-t\hat{\mathbf{Z}}+\tilde{\hh}-
t\tilde{\mathbf{Z}})\X\right)\geq (t-1)N_0)\right).
\end{align}
Let us define $y=tr\left((\hat{\hh}-t\hat{\mathbf{Z}}+\tilde{\hh}-
t\tilde{\mathbf{Z}}) \X \right)$. For given $\X$, $\hat{\hh}$, and
$\hat{\mathbf{Z}}$, we know from the results of \cite{Chalise} that
$y$ is a Gaussian distributed random variable with mean
$\mu=tr\left((\hat{\hh}-t\hat{\mathbf{Z}})\X\right)$ and
variance $\sigma_y^2=(\sigma_{\tilde{H}}^2+t^2\sigma_{\tilde{Z}}^2)\,tr(\X\X^\dagger)$.
Then, the non-outage probability  can be written as
\begin{align}
Pr(y\geq (t-1)N_0) &=\int_{ (t-1)N_0}^\infty
\frac{1}{\sqrt{2\pi}\sigma_y}\exp\left(-\frac{(y-\mu)^2}{2\sigma_y^2}\right)\\
&=\frac{1}{2}-\frac{1}{2}\text{erf}\left(\frac{(t-1)N_0-\mu}{\sqrt{2}\sigma_y}\right)\geq
\varepsilon,
\end{align}
or equivalently as,
\begin{align}
\frac{(t-1)N_0-\mu}{\sqrt{2}\sigma_y}\leq
\text{erf}^{-1}(-2\varepsilon+1).
\end{align}
Note that $\varepsilon$ should be close to one for good
performance. Thus, both $-2\varepsilon+1$ and
$\frac{(t-1)N_0-\mu}{\sqrt{2}\sigma_y}$ should be negative valued.
Note further that we have $tr(\X\X^\dagger)=\|\X\|^2$, and hence
$\sigma_y=\sqrt{\sigma_{\tilde{H}}^2+t^2\sigma_{\tilde{Z}}^2}\|\X\|$. Then, this
constraint can be written as
\begin{align}
\|\X\| \leq \frac{(t-1)N_0-\mu}{\sqrt{2
(\sigma_{\tilde{H}}^2+t^2\sigma_{\tilde{Z}}^2)}\text{erf}^{-1}(-2\varepsilon+1)}.
\end{align}
As a result, the optimization problem becomes
\begin{align}
\begin{split}
&\max_{\X, t} ~~~t \label{robust df1} \\
&s.t ~~||\X|| \leq \frac{(t-1)N_0-\mu}{\sqrt{2
(\sigma_{\tilde{H}}^2+t^2\sigma_{\tilde{Z}}^2)}\text{erf}^{-1}(-2\varepsilon+1)},\\
&and ~~diag(\X)\leq \mathbf{p} ( or ~~tr(\X)\leq P_T), ~~~ and ~~~\X
\succeq 0.
\end{split}
\end{align}
Using the same bisection search, we can solve this optimization
numerically.

\section{Numerical Results}\label{sec:simulation}
 We assume  that
$\{g_m\}$, $\{h_m\}, \{z_m\}$ are complex, circularly symmetric
Gaussian random variables with zero mean and variances $\sigma_g^2$,
$\sigma_h^2$, and $\sigma_z^2$ respectively. We first provide
numerical results for decode-and-forward beamforming schemes. In our
numerical results, we focus on the performance of second-hop secrecy
rate since the main emphasis of this paper is on the design  of the
beamforming system in the second-hop. Moreover, each figure is
plotted for fixed realizations of the Gaussian channel coefficients.
Hence, the secrecy rates in the plots are instantaneous secrecy
rates.

\begin{figure}
\begin{center}
\includegraphics[width = 0.7\textwidth]{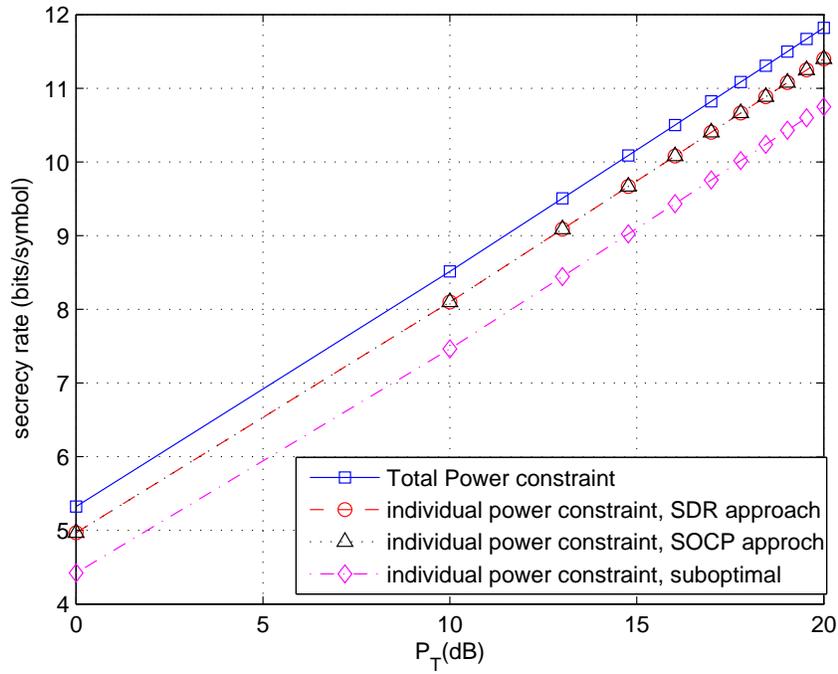}
\caption{DF Second-hop secrecy rate vs. the total relay transmit power
$P_T$ for different cases. Eavesdropper has a weaker channel.}
\label{fig:indviDF}
\end{center}
\end{figure}

\begin{figure}
\begin{center}
\includegraphics[width = 0.7\textwidth]{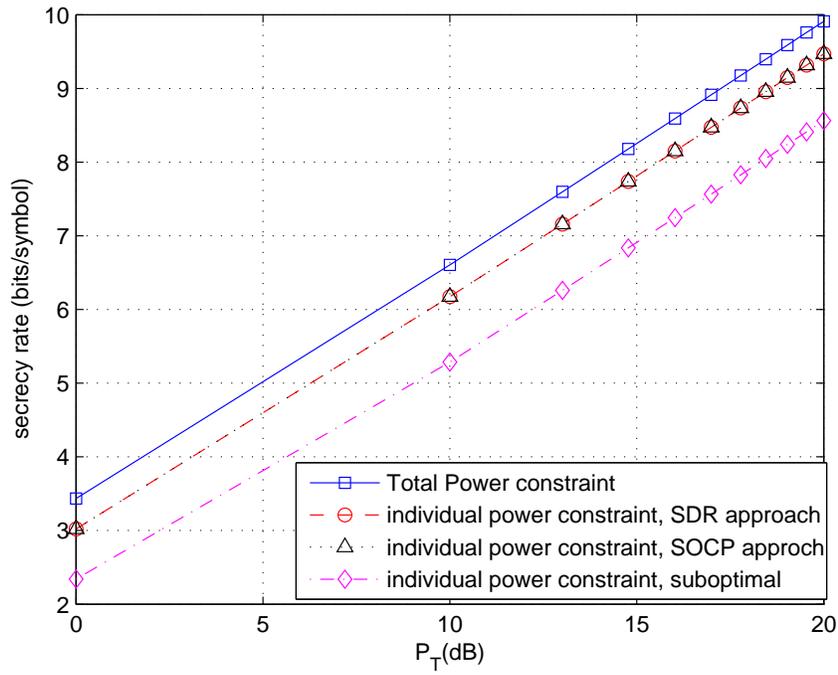}
\caption{DF Second-hop secrecy rate vs. the total relay  transmit power
$P_T$ for different cases. Eavesdropper has a stronger channel. }
\label{fig:indviDF1}
\end{center}
\end{figure}

In Figures \ref{fig:indviDF}  and \ref{fig:indviDF1}, we plot the second-hop secrecy rate, which is the maximum secrecy rate that our
collaborative relay beamforming system can support under both total
and individual relay power constraints. For the case of individual relay power
constraints, we assume that the relays have the same power budgets:
$p_i=\frac{P_T}{M}$. Specifically, in  Fig. \ref{fig:indviDF}, we
have $\sigma_h=3$, $\sigma_z=1$, $N_0=1$ and $M=5$. In this case,
the legitimate user has a stronger channel. In
Fig. \ref{fig:indviDF1}, the only changes are $\sigma_h=1$ and $\sigma_z=2$,
which imply that the eavesdropper has a stronger channel. Our CRB system can
achieve secure transmission even when the eavesdropper has more favorable channel conditions. As can be seen from
the figures, the highest secrecy rate is achieved, as expected, under a total transmit power constraint. On the other hand, we observe that only a relatively small rate loss is experienced under individual relay power constraints. Moreover, we note that our
two different optimization approaches give nearly the same result. It
also can be seen that under individual power constraint, the simple
suboptimal method suffers a constant loss as compared to SDR or SOCP
based optimal value.

\begin{figure}
\begin{center}
\includegraphics[width = 0.7\textwidth]{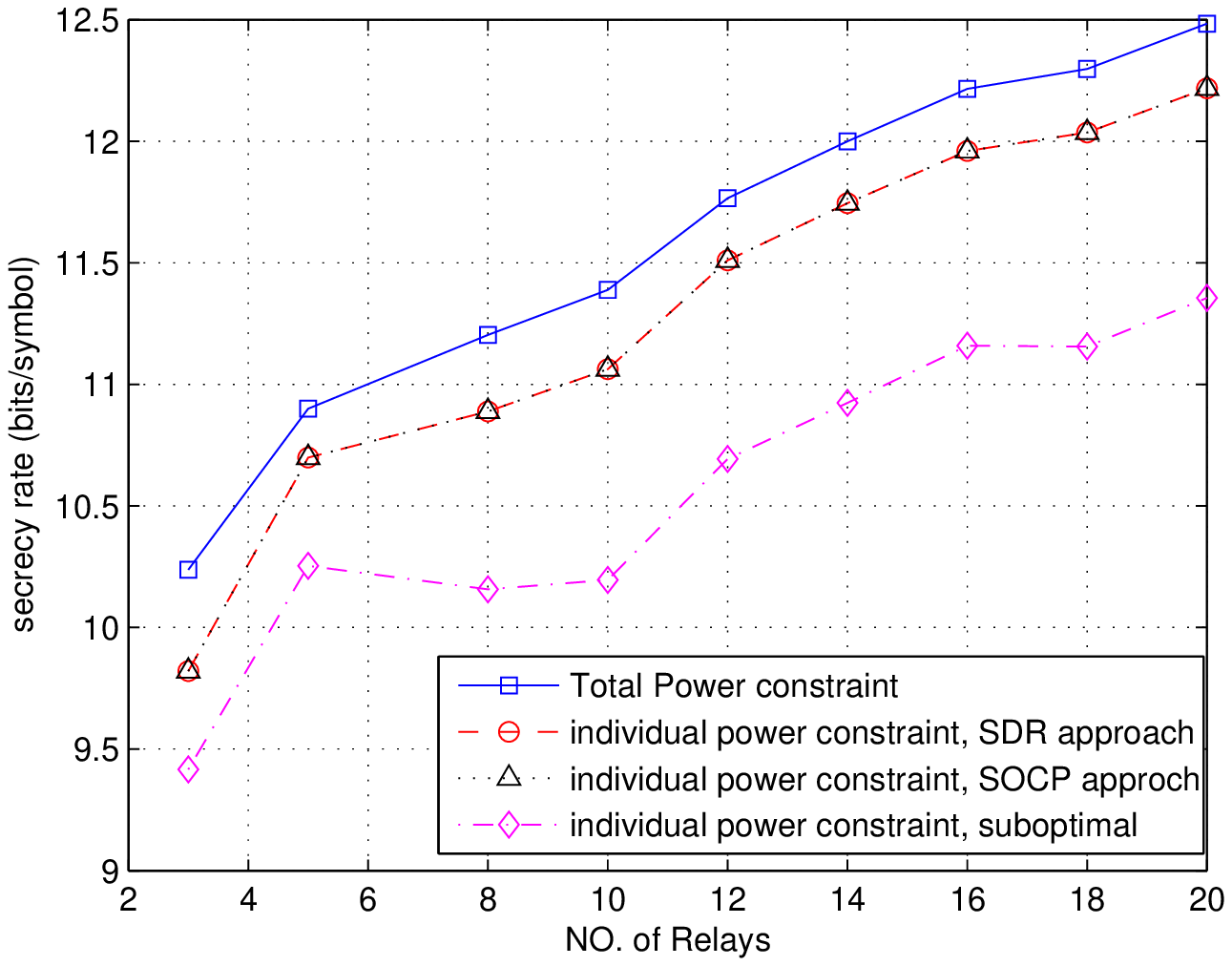}
\caption{DF second-hop secrecy rate vs. number of relays for different
cases.} \label{fig:indviDFm}
\end{center}
\end{figure}

In Fig. \ref{fig:indviDFm}, we fix the relay total transmitting
power as $P_T=10dB$, and vary the number of collaborative relays.
Other parameters are the same as those used in Fig.
\ref{fig:indviDF1}. We can see that increasing  $M$, increases the
secrecy rate under both total and individual power constraints. We
also observe that in some cases, increasing $M$ can degrade the
performance when our simplified suboptimal beamformer is used.

\begin{figure}
\begin{center}
\includegraphics[width = 0.7\textwidth]{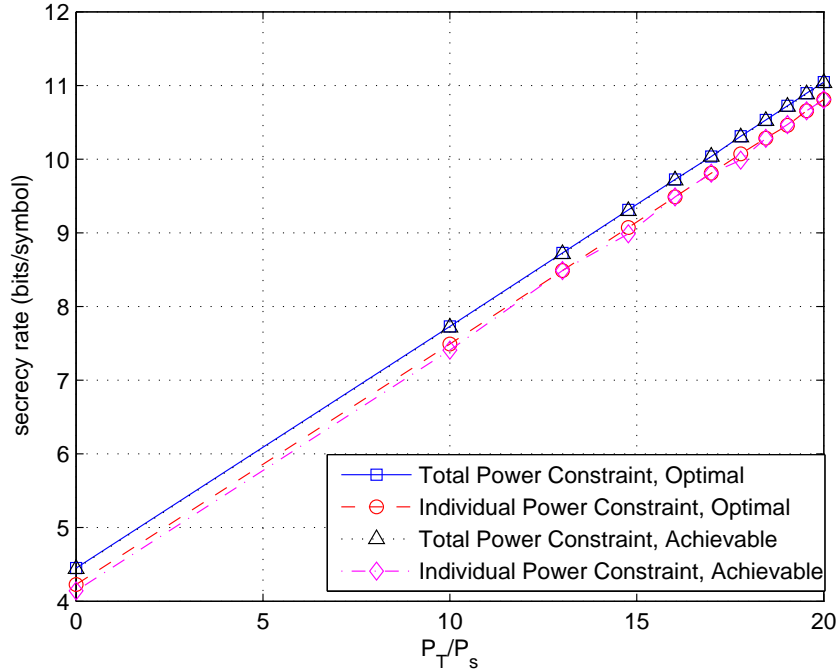}
\caption{AF secrecy rate vs. $P_T/P_s$. $\sigma_g= 10, \sigma_h=2,
\sigma_z=2, M=10$. } \label{fig:AF1}
\end{center}
\end{figure}

In Fig. \ref{fig:AF1}, we plot the secrecy rate for
amplify-and-forward collaborative relay beamforming system for both
individual and total power constraints. We also provide the result
of suboptimal achievable secrecy rate for comparison. The fixed parameters are $\sigma_g= 10, \sigma_h=2,
\sigma_z=2$, and $M=10$. Since the AF secrecy rates depend on both the source and relay powers, the rate curves are plotted as a function of $P_T/P_s$. As before, we assume that the relays have equal powers in the case in which individual power constraints are imposed, i.e., $p_i = P_T/M$.  It is immediately seen
from the figure that the achievable rates for both total and individual power constraints are very close to the
corresponding optimal ones. Thus, the achievable beamforming scheme is a good
alternative in the amplify-and-forward relaying case due to the fact that it has much less computational burden. Moreover, we interestingly observe that imposing individual relay power constraints leads to only small losses in the secrecy rates with respect to the case in which we have total relay power constraints.

\begin{figure}
\begin{center}
\includegraphics[width = 0.7\textwidth]{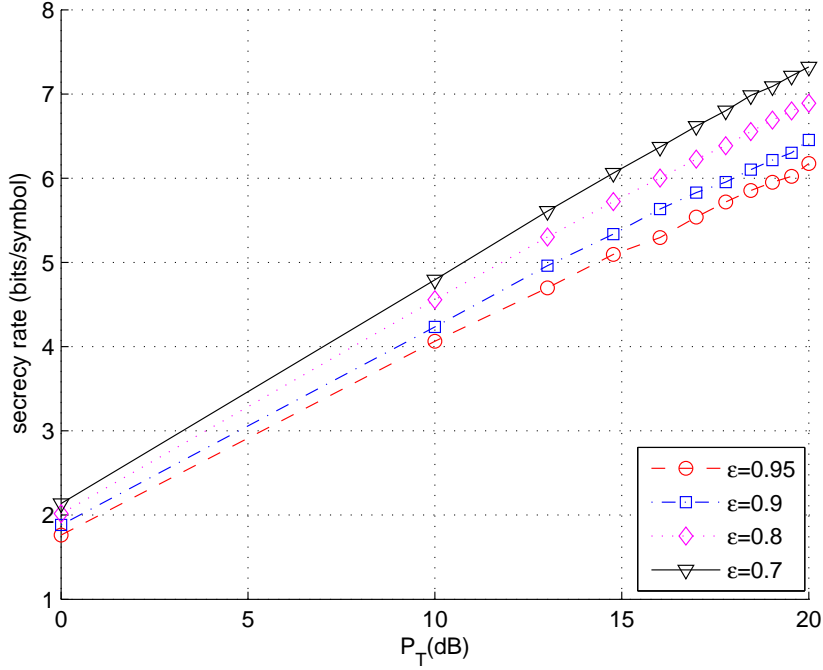}
\caption{DF second secrecy rate vs. $P_T$ under different
$\varepsilon$.} \label{fig:DFroubst}
\end{center}
\end{figure}

In Fig. \ref{fig:DFroubst},  we plot the maximum second hop secrecy
rate of decode-and-forward that we can achieve for different power
$P_T$ and non-outage probability $\varepsilon$ values. In this figure,
we fix $M=5$. $\hat{\h}$ and $\hat{\z}$ are randomly picked from
Rayleigh fading with $\sigma_{\hat{h}}=1$ and $\sigma_{\hat{z}}=2$, and we assume that
estimation errors are inversely proportional to $P_T$. More specifically, in
our simulation, we have $\sigma_{\tilde{H}}^2=0.1/P_T$ and $\sigma_{\tilde{Z}}^2=0.2/P_T$. We also
assume the relays are operating under equal individual power constraints, i.e.,
$p_i=\frac{P_T}{M}$. It is immediately observed in Fig. \ref{fig:DFroubst} that
smaller rates are supported under higher non-outage probability requirements.
In particular, this figure illustrates that our formulation and the proposed
optimization framework can be used to determine how much secrecy rate can be supported at what percentage of the time.
 For instance, at $P_T = 20dB$, we see that approximately 7.4 bits/symbol secrecy rate can be attained 70 percent of the time (i.e., $\varepsilon = 0.7$) while supported secrecy rate drops to about 6.2 bits/symbol when $\varepsilon = 0.95$.

\section{Conclusion}\label{sec:con}
In this paper, collaborative beamforming for both DF and AF
relaying is studied under secrecy constraints. Optimal beamforming designs that maximize secrecy rates
are provided under both total and individual relay power
constraints. For DF with total power constraint, we have remarked that the optimal
beamforming vector is the solution of a Rayleigh quotient problem. We have further identified the beamforming structure in the high-
and low-SNR regimes. For DF with individual relay power constraints and AF
with both total and individual relay power constraints, we have formulated the problem as a
semidefinite programming problem and provided an optimization framework.
We have also provided an alternative SOCP method to solve the DF relaying
with individual power constraints. In addition, for DF relaying, we have described the worst-case robust beamforming design  when CSI is
imperfect but bounded, and the statistical robust beamforming design based upon
minimum non-outage probability criterion. Finally, we have provided numerical results to illustrate the performance of beamforming techniques under different assumptions, e.g., DF and AF relaying, total and individual relay power constraints, perfect and imperfect channel information.

\end{spacing}

\end{document}